\pdfoutput=1
\RequirePackage{ifpdf}
\ifpdf 
\documentclass[pdftex]{sigma}
\else
\documentclass{sigma}
\fi

\def \b#1{\overline{#1}}
\def \t#1{\widetilde{#1}}
\def \h#1{\widehat{#1}}

\usepackage{bm}

\numberwithin{equation}{section}

\newtheorem{Theorem}{Theorem}[section]
\newtheorem{Proposition}[Theorem]{Proposition}

\begin{document}


\renewcommand{\PaperNumber}{036}

\FirstPageHeading

\ShortArticleName{Darboux and Binary Darboux Transformations for Discrete Integrable Systems.~II}

\ArticleName{Darboux and Binary Darboux Transformations\\ for Discrete Integrable Systems.\\
II.~Discrete Potential mKdV Equation}

\Author{Ying SHI~$^\dag$, Jonathan NIMMO~$^\ddag$ and Junxiao ZHAO~$^\S$}

\AuthorNameForHeading{Y.~Shi, J.~Nimmo and J.X.~Zhao}

\Address{$^\dag$~School of Science, Zhejiang University of Science and Technology,\\
\hphantom{$^\dag$}~Hangzhou 310023, P.R.~China}
\EmailD{\href{mailto:yingshi@zust.edu.cn}{yingshi@zust.edu.cn}}

\Address{$^\ddag$~Department of Mathematics, University of Glasgow, Glasgow G12 8QQ, UK}
\EmailD{\href{mailto:jonathan.nimmo@glasgow.ac.uk}{jonathan.nimmo@glasgow.ac.uk}}

\Address{$^\S$~School of Mathematics, University of Chinese Academy of Sciences,\\
\hphantom{$^\S$}~Beijing 100190, P.R.~China}
\EmailD{\href{mailto: jxzhao@gucas.ac.cn}{jxzhao@gucas.ac.cn}}

\ArticleDates{Received December 17, 2016, in f\/inal form May 16, 2017; Published online May 28, 2017}

\Abstract{The paper presents two results. First it is shown how the discrete potential mo\-dif\/ied KdV equation and its Lax pairs in matrix form arise from the Hirota--Miwa equation by a 2-periodic reduction. Then Darboux transformations and binary Darboux transformations are derived for the discrete potential modif\/ied KdV equation and it is shown how these may be used to construct exact solutions.}


\Classification{39A14; 37K10; 35C08}

\section{Introduction}
The discrete version of the potential modif\/ied KdV equation that we want to investigate in this paper is the nonlinear partial dif\/ference equation
\begin{gather}\label{d-p-mKdV-1}
Q(v,v_1,v_2,v_{12};a_1,a_2)\equiv a_1(vv_2-v_{1}v_{12})=a_2(vv_1-v_2v_{12}).
\end{gather}
The notation we adopted here and later is as follows, with forward shift operators $T_{n_1}$, $T_{n_2}$:
\begin{alignat*}{3}
& v:=v(n_1, n_2),\qquad & &v_1:=T_{n_1}(v)=v(n_1+1, n_2),& \\
& v_2:=T_{n_2}(v)=v(n_1, n_2+1),\qquad & &v_{12}:=T_{n_1}T_{n_2}(v)=v(n_1+1, n_2+1),&
\end{alignat*}
and $a_1$, $a_2$ denote lattice parameters associated with the directions $n_1$, $n_2$ respectively. Equation~\eqref{d-p-mKdV-1} was derived in \cite{Nijhoff-2009} from the Cauchy matrix approach, and was originally found in~\cite{Nijhoff-1983, Nijhoff-1984} through the direct linearization approach. Up to a gauge transformation $v\rightarrow i^{n_1+n_2}v$ and changing the lattice parameters as their reciprocals, equation~\eqref{d-p-mKdV-1} is equivalent to the equation H3$_{\delta=0}$ in the Adler--Bobenko--Suris (ABS) classif\/ication~\cite{ABS-2002},
\begin{gather}\label{H3}
H3_\delta\equiv a_1(vv_1+v_{2}v_{12})-a_2(vv_2+v_1v_{12})=\delta\big(a_2^2-a_1^2\big).
\end{gather}
There are several papers dedicated to closed-form $N$-soliton solutions of the `ABS list' \cite{Bulter-2012, Nalini-2010, Hietarinta-2009, Nijhoff-2009}. So, in~\cite{Nijhoff-2009}, based on a Cauchy matrix structure, the closed-form $N$-soliton solution of equation~\eqref{d-p-mKdV-1} was derived, in~\cite{Hietarinta-2009}, following Hirota's method, the authors derive bilinear dif\/ference equations of equation~\eqref{H3} and its $N$-soliton solutions in terms of Casoratian determinants, and in~\cite{Bulter-2012}, by the discrete inverse scattering transform, the authors point out that the soliton solutions of equation~\eqref{H3} derived from the Cauchy matrix approach are exactly the solutions obtained from ref\/lectionless potentials.

The Hirota--Miwa equation \cite{Hirota-1981, Miwa-1982} is the three-dimensional discrete integrable system
\begin{gather}\label{H-M-1}
(a_1-a_2)\tau_{12}\tau_3+(a_2-a_3)\tau_{23}\tau_1+(a_3-a_1)\tau_{13}\tau_2=0,
\end{gather}
where lattice parameters $a_k$ are constants, $k=1, 2, 3$, and for $\tau=\tau(n_1,n_2,n_3)$ each subscript $i$ denotes a forward shift in the corresponding discrete variable $n_i$. It was discovered by Hirota~\cite{Hirota-1981} as a fully discrete analogue of the two-dimensional Toda equation and later Miwa~\cite{Miwa-1982} showed that it was intimately related to the KP (Kadomtsev--Petviashvili) hierarchy. In paper~\cite{Hirota-1998}, Hirota gives the discretization of the potential modif\/ied KdV equation, which can be transformed into the form~\eqref{d-p-mKdV-1}, and shows that it is a 4-reduction of the Hirota--Miwa equation (which Hirota named as the discrete analogue of a generalized Toda equation).

In this paper, we discuss in detail the Darboux and binary Darboux transformations and how these may be used to obtain exact solutions of the discrete potential modif\/ied Korteweg--de~Vries (d-p-mKdV) equation \eqref{d-p-mKdV-1}. In contrast to the approaches presented in \cite{BS-2008,Nijhoff-2009,Nijhoff-1983}, we get \eqref{d-p-mKdV-1} and its Lax pairs by reducing the Hirota--Miwa equation~\eqref{H-M-1} and its Lax pairs. In fact, the 2-periodic reduction method studied here has already been investigated in \cite{ALN-2012} where authors present a multidimensionally consistent hierarchy of discrete systems whose f\/irst member is the equation \eqref{d-p-mKdV-1}. Otherwise, this was ref\/ined and extended to the non-commutative case in~\cite{D-2013}. In \cite{ALN-2012, BS-2002, BS-2008, D-2013, FN-JPA-2001}, as we see that the integrability is understood in the sense of the multidimensional consistency property, which gives a Lax pair directly. We here, through a~2-periodic reduction of the linear systems of the Hirota--Miwa equation~\eqref{H-M-1}, obtain the Lax pairs of the equation~\eqref{d-p-mKdV-1} which allows the application of the classical Darboux transforma\-tions~\mbox{\cite{Matveev-1979, Matveev-1991}}. However, up to gauge transformations, these Lax pairs are coincident with the ones given by the multidimensional consistency property~\cite{BS-2008}. This paper is part of the work which will explore the equations in the ABS list, their Lax pairs and Darboux transformations as reductions of the Hirota--Miwa equation.

The outline of this paper is as follows. In Section~\ref{Hirota--Miwa}, we recall important results on Darboux transformations and binary Darboux transformations of the Hirota--Miwa equation. In particular, in a departure from the results in \cite{Nimmo-1997, Nimmo-Chaos-2000, Shi-2014}, we write the linear system of Hirota--Miwa equation in a dif\/ferent form, by the gauge transformation $\phi\rightarrow \prod\limits_{i=1}^{3}a_i^{-n_i}\phi$, which is suitable for making the reduction. In Section~\ref{sec-3}, we show that how the d-p-mKdV equation and its Lax pairs in matrix form arise from the Hirota--Miwa equation by a $2$-periodic reduction. Then its Darboux transformations and binary Darboux transformations are derived and it is shown how these may be used to construct exact solutions.

\section{Hirota--Miwa equation}\label{Hirota--Miwa}
The Hirota--Miwa equation \eqref{H-M-1} arises as the compatibility conditions of the linear system
\begin{gather}\label{H-M-LP-1}
\phi_i-\phi_j=(a_i-a_j)u^{ij}\phi, \qquad 1\leq i<j\leq3,
\end{gather}
where for $\phi=\phi(n_1,n_2,n_3)$ each subscript $i$ denotes a forward shift in the corresponding discrete variable~$n_i$, for example, $\phi_{1} = T_{_{n_1}} (\phi) = \phi(n_1+1,n_2,n_3)$. This linear system~\eqref{H-M-LP-1} is compatible if and only if
\begin{subequations}\label{dKP-u-form}
\begin{gather}
(a_1-a_2)u^{12}+(a_2-a_3)u^{23}+(a_3-a_1)u^{13}= 0, \label{dKP-u-form-a}\\
\big(u^{ij}\big)_{_{k}} u^{ik}=\big(u^{ik}\big)_{_{j}} u^{ij}. \label{dKP-u-form-b}
\end{gather}
\end{subequations}
Note that when one uses the formula $u^{ij}=\tau_{ij}\tau/\tau_i\tau_j$, \eqref{dKP-u-form-a} gives~\eqref{H-M-1} and~\eqref{dKP-u-form-b} is satisf\/ied identically. A second way is to suppose $u^{ij}=(v_j-v_i+(a_i-a_j))/(a_i-a_j)$. This ansatz solves~\eqref{dKP-u-form-a} exactly and~\eqref{dKP-u-form-b} becomes the discrete potential KP (d-p-KP) equation~\cite{Nijhoff-1984-dpkp}.

In this paper, in particular we deal with the Hirota--Miwa equation~\eqref{H-M-1} together with its linear system in the form~\eqref{H-M-LP-1}. Using the reversal-invariance property of the Hirota--Miwa equation, i.e., it is invariant with respect to the reversal of all lattice directions $n_i\rightarrow -n_i$, we have the linear system in formal adjoint form~\cite{Nimmo-1997}
\begin{gather}\label{H-M-LP-2}
\psi_{\b i}-\psi_{\b j}=(a_i-a_j)\frac{\tau_{\b i\b j}\tau}{\tau_{\b i}\tau_{\b j}}\psi,\qquad 1\leq i<j\leq3.
\end{gather}
The subscript $\b i$ denotes a backward shift with respect to $n_i$, for example, $\psi_{\b 1} = T_{_{n_1}}^{-1}(\psi) = \psi(n_1-1,n_2,n_3)$.

\subsection{Darboux and binary Darboux transformations}

The basic Darboux transformation for the Hirota--Miwa equation is stated in the following proposition.
\begin{Proposition}\label{prop1-HM}
Let $\theta$ be a non-zero solution of the linear system \eqref{H-M-LP-1} for some~$\tau$. Then the transformation
\begin{gather*}
\mathrm{DT}^{\theta}\colon \ \phi\rightarrow\frac{C _{_{[i]}} (\theta, \phi)}{\theta}, \qquad \tau\rightarrow\theta\tau,
\end{gather*}
leaves \eqref{H-M-LP-1} invariant, where $C _{_{[i]}} (\theta, \phi)=\theta\phi_i-\theta_i\phi$, $i=1,2,3$, using the subscript~$[i]$ to designate that the forward shifts of the determinant $C _{_{[i]}} (\theta, \phi)$ is with respect to the variable~$n_i$.
\end{Proposition}

Next we write down the closed form expression for the result of~$N$ applications of the above Darboux transformation, which give solutions in Casoratian determinant form. To do this we need to def\/ine the Casoratian of $N$ solutions. Let $\bm\theta=({\theta^{1}}(n_1,n_2,n_3), {\theta^{2}}(n_1,n_2,n_3), \dots$, ${\theta^{N}}(n_1,n_2, n_3))^T$ be an $N$-vector solution of~\eqref{H-M-LP-1}. The Casoratian determinant (with forward-shifts) can be written as
\begin{gather*}
C\big(\theta^{1},\theta^{2}, \dots, \theta^{N}\big) = \big|\bm\theta,T_{_{n_i}}(\bm\theta), T_{_{n_i}}^2(\bm\theta)\dots, T_{_{n_i}}^{N-1}(\bm\theta)\big|, \qquad 1\leq i \leq 3,
\end{gather*}
which may also be unambiguously def\/ined in the following notation as
\begin{gather*}
C_{_{[i]}}\big(\theta^{1},\theta^{2}, \dots, \theta^{N}\big) = \big|\bm\theta(0),\bm\theta(1), \bm\theta(2), \dots, \bm\theta(N-1)\big|, \qquad 1\leq i \leq 3,
\end{gather*}
where $\bm\theta(k)$ denotes the $N$-vector $\big({\theta^{1}}(n_1,n_2,n_3), {\theta^{2}}(n_1,n_2,n_3), \dots, {\theta^{N}}(n_1,n_2,n_3)\big)^T$ subject to the~$k$ times shift $T_{_{n_i}}^{k}$ on $n_i$ which gives $n_i \rightarrow n_i+k$, $0\leq k \leq N-1$, and $i=1, 2$ or $3$, the same value being taken for~$i$ in each column in the determinant. Then we have the following.

\begin{Proposition}\label{prop1N-HM}
Let $\theta^{1},\theta^{2}, \dots, \theta^{N}$ be non-zero, independent solutions of the linear system~\eqref{H-M-LP-1} for some~$\tau$, such that $C_{_{[i]}}\big(\theta^{1},\theta^{2}, \dots, \theta^{N}\big)\neq0$. Then the $N$-fold Darboux transformation
\begin{gather*}
\phi\rightarrow\frac{C_{_{[i]}}\big(\theta^{1},\theta^{2}, \dots, \theta^{N},\phi\big)}{C_{_{[i]}}\big(\theta^{1},\theta^{2}, \dots, \theta^{N}\big)}, \qquad \tau\rightarrow C_{_{[i]}}\big(\theta^{1},\theta^{2}, \dots, \theta^{N}\big)\tau,
\end{gather*}
leaves \eqref{H-M-LP-1} invariant.
\end{Proposition}

Now we can apply the ref\/lections $n_i \rightarrow -n_i$, $i=1, 2, 3$, to the above results to deduce adjoint Darboux transformation for the second linear system~\eqref{H-M-LP-2}.
\begin{Proposition}\label{prop2-HM}
Let $\rho$ be a non-zero solution of the linear system \eqref{H-M-LP-2} for some $\tau$. Then the transformation
\begin{gather*}
\mathrm{DT}^{\rho}\colon \ \psi\rightarrow\frac{C _{_{[\b i]}} (\rho, \psi)}{\rho}, \qquad \tau\rightarrow\rho\tau,
\end{gather*}
leaves \eqref{H-M-LP-2} invariant, where $C _{_{[\b i]}} (\rho, \psi)=\rho_{_{\b i}}\psi-\rho\psi_{_{\b i}}$, $i=1,2,3$, using the subscript $[\b i]$ to designate that the backward shifts of the determinant $C _{_{[\b i]}} (\rho, \psi)$ is with respect to the variable~$n_i$.
\end{Proposition}

The $N$-fold adjoint Darboux transformation is expressed in terms of the Casoratian
\begin{gather*}
C_{_{[\b i]}}\big(\rho^{1},\rho^{2}, \dots, \rho^{N}\big) = |\bm\rho(0),\bm\rho(-1), \bm\rho(-2), \dots, \bm\rho(-N+1)|, \qquad 1\leq i \leq 3,
\end{gather*}
where $\bm \rho=\big(\rho^1, \rho^2, \dots, \rho^N\big)^T$ and $\bm\rho(-k)=T_{_{n_i}}^{-k}(\bm \rho)=\bm\rho|_{_{n_i\rightarrow n_i-k}}$, $0\leq k\leq N-1$, the same $i=1, 2$ or~$3$ be taken in all columns.

\begin{Proposition}\label{prop2N-HM}
Let ${\rho^{1}}, {\rho^{2}}, \dots, {\rho^{N}}$ be $N$ non-zero independent solutions of the linear system~\eqref{H-M-LP-2} for some $\tau$, such that $C_{_{[\b i]}}\big(\rho^{1},\rho^{2}, \dots, \rho^{N}\big)\neq 0$. Then the $N$-fold adjoint Darboux transformation
\begin{gather*}
\psi\rightarrow\frac{C_{_{[\b i]}}\big(\rho^{1},\rho^{2}, \dots, \rho^{N},\psi\big)}{C_{_{[\b i]}}\big(\rho^{1},\rho^{2}, \dots, \rho^{N}\big)},\qquad \tau\rightarrow C_{_{[\b i]}}\big(\rho^{1},\rho^{2}, \dots, \rho^{N}\big)\tau,
\end{gather*}
leaves \eqref{H-M-LP-2} invariant.
\end{Proposition}

To construct a binary Darboux transformation, we introduce the potential $\omega=\omega(\phi, \psi)$, def\/ined by the relations
\begin{gather}
\Delta_i\omega(\phi, \psi) =\phi \psi_i, \qquad i=1, 2, 3, \label{omega-1}
\end{gather}
where $\Delta_i=T_{n_i}-1$ is the forward-dif\/ference operator in discrete variable~$n_i$. If~$\phi$ and~$\psi$ satisfy the linear systems~\eqref{H-M-LP-1} and~\eqref{H-M-LP-2} for some~$\tau$, respectively, then~\eqref{omega-1} are compatible in the sense as $\Delta_{i}(\phi \psi_j)=\Delta_{j}(\phi \psi_i)$, for $i<j$. So the potential $\omega$ is well-def\/ined.

The following proposition gives the binary Darboux transformation of the Hirota--Miwa equation~\eqref{H-M-1}.

\begin{Proposition}\label{prop12-HM}
For some $\tau$, let $\theta$ and $\phi$ be two non-zero solutions of the linear system~\eqref{H-M-LP-1}, $\rho$ and $\psi$ be two non-zero solutions of the linear system \eqref{H-M-LP-2}, then
\begin{alignat*}{4}
& \mathrm{BDT^{\theta, \rho}}\colon \quad && \phi\rightarrow\phi-\theta \omega(\theta, \rho)^{-1}\omega(\phi,\rho), \qquad && \tau\rightarrow\omega(\theta, \rho)\tau,& \\
& \mathrm{aBDT^{\theta, \rho}}\colon \quad && \psi\rightarrow\psi-\rho \omega(\theta, \rho)^{-1}\omega(\theta,\psi), \qquad && \tau\rightarrow\omega(\theta, \rho)\tau, & 
\end{alignat*}
leave \eqref{H-M-LP-1} and \eqref{H-M-LP-2} respectively invariant.
\end{Proposition}

The $N$-fold iteration of these binary Darboux transformations are given below.

\begin{Proposition}\label{prop12N-HM}
Let $\bm\theta=\big(\theta^1,\dots,\theta^N\big)^T$ and $\bm\rho=\big(\rho^1,\dots,\rho^N\big)^T$ satisfy linear systems~\eqref{H-M-LP-1} and~\eqref{H-M-LP-2} for some~$\tau$ respectively. Then
\begin{alignat*}{3}
& \phi\rightarrow\begin{vmatrix}
\omega\big(\bm\theta, \bm \rho^T\big) & \bm\theta\\
\omega\big(\phi, \bm \rho^T\big) & \phi \\
\end{vmatrix}
\big|\omega\big(\bm\theta, \bm \rho^T\big) \big|^{-1},
\qquad &&
\tau\rightarrow\big|\omega\big(\bm\theta, \bm \rho^T\big)\big|\tau,&
\\ 
& \psi\rightarrow \begin{vmatrix}
\omega\big(\bm\theta^T, \bm \rho\big) &\bm \rho \\
\omega\big( \bm\theta^T,\psi\big) & \psi \\
\end{vmatrix}
\big|\omega\big(\bm\theta^T, \bm \rho\big) \big|^{-1},\qquad&&
\tau\rightarrow\big|\omega\big(\bm\theta^T, \bm \rho\big)\big|\tau,&
\end{alignat*}
leave \eqref{H-M-LP-1} and \eqref{H-M-LP-2} respectively invariant.

Here $\omega\big(\bm\theta, \bm \rho^T\big)=\big(\omega\big(\theta^{(i)}, \rho^{(j)}\big)\big)_{i,j=1,\dots,N}$, $\omega\big(\bm\theta^T, \bm\rho\big)=\omega\big(\bm\theta, \bm \rho^T\big)^T$ are $N\times N$ matrix, $\omega\big(\phi, \bm \rho^T\big)=
\big(\omega\big(\phi,\rho^{(j)}\big)\big)_{j=1,\dots,N}$ and $\omega\big(\bm \theta^T,\psi\big)=\big(\omega\big(\theta^{(i)},\psi\big)\big)_{i=1,\dots,N}$ are $N$-row vectors.
\end{Proposition}

The proofs of those above propositions are straightforward computation, so we do not give the details. The reader is also referred to the papers \cite{Nimmo-1997, Nimmo-Chaos-2000}.

\subsection{Explicit solutions obtained by Darboux transformations}
Here we present explicit examples of the classes of solutions that may be obtained by means of the Darboux transformations derived above. We choose the seed solution of the Hirota--Miwa equation \eqref{H-M-1} as $\tau=\tau_0=1$. With this choice, the f\/irst linear system \eqref{H-M-LP-1} reads
\begin{gather*}
\phi_i-\phi_j=(a_i-a_j)\phi, \qquad 1\leq i<j\leq3,
\end{gather*}
and the basic eigenfunctions, depending on a single parameter $p$ are found to be
\begin{gather}\label{seed-linear-solu-1}
\phi(n_1, n_2, n_3; p)=\prod_{i=1}^{3}(a_i+p)^{n_i}.
\end{gather}
In a similar way the basic eigenfunctions of the adjoint linear system~\eqref{H-M-LP-2}, depending on a single parameter $q$, are
 \begin{gather}\label{seed-linear-solu-2}
\psi(n_1, n_2, n_3; q)=\prod_{i=1}^{3}(a_i+q)^{-n_i}.
\end{gather}
For these eigenfunctions above we may integrate \eqref{omega-1} and obtain the potential
\begin{gather}\label{seed-linear-solu-omeg}
\omega(\phi, \psi)=\frac{1}{p-q}\prod_{i=1}^{3}\left(\frac{a_i+p}{a_i+q}\right)^{n_i}+c.
\end{gather}
Given the above expression it is straightforward to write down the following explicit solution for the Hirota--Miwa equation~\eqref{H-M-1}
\begin{gather}\label{n-solu-1}
\tau(n_1, n_2, n_3)=C_{{_{[i]}}}\big(\theta^1, \theta^2, \dots, \theta^N\big)\tau_0,
\end{gather}
where $\theta^k=\alpha_k\theta(n_1, n_2, n_3; p_k)+\theta(n_1, n_2, n_3; p^\prime_k)$ where $\theta(n_1,n_2,n_3; p_k)$ and $\theta(n_1,n_2,n_3; p_k^{\prime})$ are given by~\eqref{seed-linear-solu-1} and~$p_k$, $p^\prime_k$, $p_k\neq p^\prime_k$ and $\alpha_k$ are arbitrary constants;
\begin{gather*}
\tau(n_1, n_2, n_3)=C_{{_{[\b i]}}}\big(\rho^1, \rho^2, \dots, \rho^N\big)\tau_0,
\end{gather*}
where $\rho^k=\beta_k\rho(n_1, n_2, n_3; q_k)+\rho(n_1, n_2, n_3; q^\prime_k)$ where $\rho(n_1,n_2,n_3; q_k)$ and $\rho(n_1,n_2,n_3; q_k^{\prime})$ are given by~\eqref{seed-linear-solu-2} and~$q_k$, $q^\prime_k$, $q_k\neq q^\prime_k$ and $\beta_k$ are arbitrary constants;
\begin{gather*}
\tau(n_1, n_2, n_3)=\det (\omega_{k,l})\tau_0 , \qquad k, l= 1, 2, \dots, N,
\end{gather*}
where $\omega_{k,l}$ is given by~ \eqref{seed-linear-solu-omeg} with and $p=p_k$, $q=q_l$ and $c=c_{kl}$.

\section{Discrete potential modif\/ied KdV equation}\label{sec-3}

\subsection[From Hirota--Miwa equation to d-p-mKdV equation: 2-periodic reductions]{From Hirota--Miwa equation to d-p-mKdV equation:\\ 2-periodic reductions}

Here, we explain the way to obtain the d-p-mKdV equation \eqref{d-p-mKdV-1} from the Hirota--Miwa equa\-tion~\eqref{H-M-1} through a 2-periodic reduction technique.

For the Hirota--Miwa equation \eqref{H-M-1}, from \eqref{n-solu-1}, it is easy to get its one soliton solution in discrete exponential function form
\begin{gather}\label{tau-solution}
 \tau(n_1,n_2,n_3)=1+\alpha\prod_{i=1}^{3}\left(\frac{a_i+p}{a_i+p^\prime}\right)^{n_i}.
\end{gather}
Introduce $f(n_1, n_2, n_3)$ and $\bar f(n_1, n_2, n_3)$ and impose a 2-periodic property on the $\tau$ func\-tion~\eqref{tau-solution} as below
\begin{gather}\label{f-condition}
 \tau=f=T_{n_3}^2(f),\qquad \b f=T_{n_3}(f).
\end{gather}
Note here that the reduction condition \eqref{f-condition} gives $a_3=0$, $p^\prime=-p$, and
\begin{gather}\label{bf-condition}
 \b f=T_{n_3}^2(\b f),\quad f=T_{n_3}(\b f).
\end{gather}
Moreover \eqref{f-condition} and~\eqref{bf-condition} indicate the symmetric property between~$f$ and~$\b f$, with respect to~$n_3$.

By applying the reduction condition \eqref{f-condition} to the Hirota--Miwa equation~\eqref{H-M-1}, together with parameter reduction $a_3=0$, we get
\begin{subequations}\label{H-M-2}
\begin{gather}
(a_1-a_2)f_{12}\b f =a_1f_{2}\b f_1-a_2f_1\b f_{2},\label{Bilinear-f-g-1}\\
(a_1-a_2)\b f_{12}f =a_1\b f_{2}f_1-a_2\b f_1f_{2}.\label{Bilinear-f-g-2}
\end{gather}
\end{subequations}
There are two ways to obtain the equation \eqref{Bilinear-f-g-2}, one way is applying the symmetric property between~$f$ and~$\b f$ to the equation~\eqref{Bilinear-f-g-1}, the another one is taking the shift operator~$T_{n_3}$ on the Hirota--Miwa equation~\eqref{H-M-1}, and using the reduction condition~\eqref{bf-condition}.

Def\/ine two functions (potentials)
\begin{gather}\label{potentials}
 v(n_1,n_2,n_3)=\frac{\b f}{f},\qquad u(n_1,n_2,n_3)=\frac{f_{12}f}{f_1f_{2}}.
\end{gather}
By substituting \eqref{potentials} into \eqref{H-M-2}, we get
\begin{subequations}\label{d-Miura}
\begin{gather}
(a_1-a_2)vu =a_1v_1-a_2v_{2},\label{d-Miura-1}\\
(a_1-a_2)v_{12}u =a_1v_{2}-a_2v_1.\label{d-Miura-2}
\end{gather}
\end{subequations}
Eliminating $u$ in \eqref{d-Miura} gives
\begin{gather}\label{d-p-mKdV}
Q(v,v_1,v_2,v_{12};a_1,a_2)\equiv v_{12}(a_1 v_{1}-a_2v_2)-v(a_1v_2-a_2v_1)=0,
\end{gather}
which is the d-p-mKdV equation \eqref{d-p-mKdV-1} and is exactly same as the one f\/irst given by Nijhof\/f, cf.~\cite{Nijhoff-2009}, through the Cauchy matrix approach. Moreover, the relation \eqref{d-Miura} serves as the discrete Miura transformation between the d-KdV equation
\begin{gather*}
\frac{1}{u_1}-\frac{1}{u_2}=\frac{a_1-a_2}{a_1+a_2}\left (u_{12}-u\right),
\end{gather*}
in potential $u$ (or more specif\/ically, say~$u_{\b 2}$~\cite{Shi-2014}) and the d-p-mKdV equation~\eqref{d-p-mKdV} in potential~$v$.

Another interesting result is that with the periodic property of $f$ and $\b f$, we have the following formulae on the potentials~$u$ and~$v$ as follows
\begin{gather*}
T_{n_3}(u)=uv_{12}v v_1^{-1} v_2^{-1},\qquad T_{n_3}^2(u)=u, \qquad T_{n_3}(v)=v^{-1}, \qquad T_{n_3}^2(v)=v.
\end{gather*}
So the potentials $u$ and $v$ also satisfy the 2-periodic property in the virtual variable~$n_3$. We observe that if $v$ is a solution of the d-p-mKdV equation then, as in the continuous case, $-v$ is a solution, but in the discrete case, $v^{-1}$ is a yet another solution.

Under the reduction condition \eqref{f-condition}, from the $\tau$ function~\eqref{tau-solution}, we easily get the exact solution of~\eqref{H-M-2}
\begin{gather*}
f(n_1,n_2,n_3)=1+\alpha(-1)^{n_3}\prod_{i=1}^{2}\left(\frac{a_i+p}{a_1-p}\right)^{n_i},\\
\b f(n_1,n_2,n_3)=1-\alpha(-1)^{n_3}\prod_{i=1}^{2}\left(\frac{a_i+p}{a_1-p}\right)^{n_i},
\end{gather*}
which directly gives the one soliton solution of the d-p-mKdV equation
\begin{gather*}
 v(n_1,n_2)=\frac{\b f}{f}=\frac{1-\alpha(-1)^{n_3} \prod\limits_{i=1}^{2}\left(\frac{a_i+p}{a_1-p}\right)^{n_i}}{1+\alpha(-1)^{n_3} \prod\limits_{i=1}^{2}\left(\frac{a_i+p}{a_1-p}\right)^{n_i}}.
\end{gather*}
Note here that in the equation \eqref{d-p-mKdV}, there is no shift depends on the discrete variable~$n_3$. So treating the~$n_3$ as a virtual variable for the d-p-mKdV equation is allowable.

Next, we show the way of discovering the linear system in matrix form of the d-p-mKdV equation~\eqref{d-p-mKdV} from the linear system of the Hirota--Miwa equation \eqref{H-M-LP-1} through the $2$-periodic reduction technique.

Introduce eigenfunctions $\phi(n_1,n_2,n_3)$ and $\b \phi(n_1, n_2, n_3)$ and impose a 2-periodic condition on the eigenfunction~$\phi(n_1,n_2,n_3)$ in the linear system~\eqref{H-M-LP-1} as below
\begin{gather}\label{phi-condition}
\phi=\lambda^{-2}T_{n_3}^2(\phi),\qquad \b\phi=\lambda^{-1} T_{n_3}(\phi),
\end{gather}
where the parameter $\lambda$ serves as the spectral parameter. From \eqref{phi-condition}, we have
\begin{gather}\label{bphi-condition}
 \b\phi=\lambda^{-2}T_{n_3}^2(\b\phi),\qquad \phi=\lambda^{-1} T_{n_3}(\b\phi).
\end{gather}
So \eqref{phi-condition} and \eqref{bphi-condition} mean the symmetric property between~$\phi$ and~$\b\phi$, with respect to~$n_3$.

By applying the reduction conditions \eqref{f-condition} and \eqref{phi-condition}, together with $a_3=0$, to the linear system~\eqref{H-M-LP-1}, we get
\begin{subequations}\label{H-M-LP2}
\begin{gather}
\phi_{1}-\phi_{2} =(a_1-a_2)\frac{f_{12}f}{f_{1}f_{2}}\phi,\label{H-M-LP2-1}\\
\phi_{2}-\lambda\b\phi =a_2\frac{\b f_{2}f}{f_{2}\b f}\phi, \label{H-M-LP2-2}\\
\lambda\b\phi-\phi_{1} =-a_1\frac{\b f_1 f}{f_1 \b f}\phi.\label{H-M-LP2-3}
\end{gather}
\end{subequations}
Then by using the symmetric property \eqref{bf-condition} and \eqref{bphi-condition} respectively between~$f$ and~$\b f$, $\phi$ and~$\b\phi$, we get
\begin{subequations}\label{H-M-LP3}
\begin{gather}
\b\phi_{1}-\b\phi_{2}=(a_1-a_2)\frac{\b f_{12}\b f}{\b f_{1}\b f_{2}}\b\phi,\label{H-M-LP3-1}\\
\b\phi_{2}-\lambda\phi=a_2\frac{f_{2}\b f}{\b f_{2}f}\b\phi, \label{H-M-LP3-2}\\
\lambda\phi-\b\phi_{1}=-a_1\frac{f_1 \b f}{\b f_1f}\b\phi.\label{H-M-LP3-3}
\end{gather}
\end{subequations}
Substituting \eqref{potentials} into \eqref{H-M-LP2} and \eqref{H-M-LP3} gives
\begin{subequations}\label{H-M-LP4}
\begin{gather}
\phi_{1}-\phi_{2} =(a_1-a_2)u\phi,\label{H-M-LP4-1}\\
\phi_{2}-\lambda\b\phi =a_2v_2v^{-1}\phi, \label{H-M-LP4-2}\\
\lambda\b\phi-\phi_{1} =-a_1v_1 v^{-1}\phi,\label{H-M-LP4-3}
\end{gather}
\end{subequations}
and
\begin{subequations}\label{H-M-LP5}
\begin{gather}
\b\phi_{1}-\b\phi_{2} =(a_1-a_2)uv_{12}v v_{1}^{-1}v_2^{-1}\b\phi,\label{H-M-LP5-1}\\
\b\phi_{2}-\lambda\phi =a_2v_2^{-1}v\b\phi, \label{H-M-LP5-2}\\
\lambda\phi-\b\phi_{1} =-a_1v_1^{-1}v\b\phi.\label{H-M-LP5-3}
\end{gather}
\end{subequations}
Through the discrete Miura transformation \eqref{d-Miura}, the equations \eqref{H-M-LP4-1} and \eqref{H-M-LP5-1} can be derived by \eqref{H-M-LP4-2} and \eqref{H-M-LP4-3}, \eqref{H-M-LP5-2} and \eqref{H-M-LP5-3}, respectively.

Def\/ining a vector eigenfunction $\bm\Phi=(\phi,\b\phi)^T$, which satisf\/ies the condition \eqref{phi-condition}, then \eqref{H-M-LP4-3} and \eqref{H-M-LP5-3}, \eqref{H-M-LP4-2} and \eqref{H-M-LP5-2}, can be respectively written in matrix form as below
\begin{subequations}\label{matrix-LP}
\begin{gather}
\bm\Phi_1=\bm L\bm\Phi,\label{matrix-LP-1}\\
\bm\Phi_2=\bm M\bm\Phi, \label{matrix-LP-2}
\end{gather}
\end{subequations}
where
\begin{gather*}
\bm L=\left(
 \begin{matrix}
 a_1v_1v^{-1} & \lambda \\
 \lambda & a_1v_1^{-1}v
 \end{matrix}
 \right)
,\qquad
\bm M=\left(
 \begin{matrix}
 a_2v_2v^{-1} & \lambda \\
 \lambda & a_2v_2^{-1}v
 \end{matrix}
 \right).
\end{gather*}
One then f\/inds that
\begin{gather*}
0=\bm\Phi_{12}-\bm\Phi_{21}=(\bm L_2\bm M-\bm M_1\bm L)\bm\Phi=\lambda Q(v,v_{1},v_{2},v_{{12}};a_1,a_2)
\left(
\begin{matrix}
 0 & -v_1^{-1}v_2^{-1} \\
v^{-1} v_{12}^{-1} & 0
\end{matrix}
\right)\bm\Phi.
\end{gather*}
So the compatibility condition of the above linear system \eqref{matrix-LP} in eigenfunction~$\bm\Phi$ is that~$v$ obeys the d-p-mKdV equation~\eqref{d-p-mKdV}.

\subsection{Darboux and binary Darboux transformations}\label{sec-3-1}

In this section, we will see that through the reduction conditions~\eqref{f-condition} and~\eqref{phi-condition}, it is easy to investigate the Darboux and binary Darboux transformations of d-p-mKdV equation.

Let $v$ be a solution of the d-p-mKdV equation \eqref{d-p-mKdV} and $\bm\Phi=(\phi, \b\phi)^T$ be a vector solution of its Lax pair~\eqref{matrix-LP}. The fundamental Darboux transformation of the d-p-mKdV equation is given as below.
\begin{Proposition}\label{prop1-dpmKdV}
Suppose $(\theta,\b\theta)^T$, which holds the $2$-periodic property $\theta = \mu^{ -2}T_{_{n_3}}^2 (\theta)$, $\b\theta = \mu^{ -1}T_{_{n_3}} (\theta)$, is a vector solution of the linear system~\eqref{matrix-LP} by taking $\lambda=\mu$ for some $v$, then
\begin{gather}\label{dpmKdV-DT-1-1}
\mathrm {DT}^{\theta,\b\theta}\colon \
\phi\rightarrow\frac{C_{_{[3]}}(\theta, \phi)} {\theta},\qquad \b\phi\rightarrow\frac{C_{_{[3]}}(\b\theta, \b\phi)} {\b\theta},\qquad v\rightarrow \frac{T_{_{n_3}}(\theta)}{\theta}v=\mu^2\frac{\b\theta}{T_{_{n_3}}(\b\theta)}v
\end{gather}
leaves \eqref{matrix-LP} invariant. Otherwise,
\begin{gather*}
\frac{C_{_{[3]}}(\b\theta, \b\phi)} {\b\theta}=\lambda^{-1}T_{_{n_3}}\left(\frac{C_{_{[3]}}(\theta, \phi)} {\theta}\right),\qquad
\frac{C_{_{[3]}}(\theta, \phi)} {\theta}=\lambda^{-1}T_{_{n_3}}\left(\frac{C_{_{[3]}}(\b\theta, \b\phi)} {\b\theta}\right).
\end{gather*}
\end{Proposition}
We remark that may also write the gauge transformation of $\bm\Phi=(\phi, \b\phi)^T$ in~\eqref{dpmKdV-DT-1-1} in matrix form as follows
\begin{gather*}
\mathrm {DT}^{\theta,\b\theta}\colon \
\bm\Phi\rightarrow\left(\begin{matrix}-\mu\theta^{-1}\b\theta & \lambda \\ \lambda& -\mu\theta{\b\theta}^{-1}\end{matrix}\right)\bm\Phi.
\end{gather*}
But for later convenience of the construction of the binary Darboux transformation, we here write in scalar form shown in~\eqref{dpmKdV-DT-1-1}.

Next we write down the closed form expression for the result of $N$ applications of the above Darboux transformation, which give solutions in Casoratian determinant form.

\begin{Proposition}\label{prop1N-dpmKdV}
Let $\big(\theta^{1}, \b\theta^{1}\big)^T , \big(\theta^{2},\b\theta^{2}\big)^T , \dots, \big(\theta^{N}, \b\theta^{N}\big)^T$, satisfying $\theta^{k} = \lambda_k^{-2}T_{_{n_3}}^2 (\theta^{k})$, $\b\theta^{k} = \lambda_k^{-1}T_{_{n_3}} (\theta^{k})$, be~$N$ non-zero independent vector solutions of the linear system~\eqref{matrix-LP} by taking $\lambda=\lambda_k$, $k=1, 2, \dots, N$, for some~$v$, such that $C_{_{[3]}}\big(\theta^{1},\theta^{2}, \dots, \theta^{N}\big)\neq0$. Then
\begin{subequations}\label{N-DT1}
\begin{gather}
 \phi\rightarrow\t\phi=\frac{C_{_{[3]}}\big(\theta^{1},\theta^{2}, \dots, \theta^{N},\phi\big)}{C_{_{[3]}}\big(\theta^{1},\theta^{2}, \dots, \theta^{N}\big)},
\qquad \b\phi\rightarrow\t{\b\phi}=\frac{C_{_{[3]}}\big(\b\theta^{1},\b\theta^{2}, \dots, \b\theta^{N},\b\phi\big)}{C_{_{[3]}}\big(\b\theta^{1},\b\theta^{2}, \dots, \b\theta^{N}\big)},\\
 v\rightarrow \t v=\frac{T_{_{n_3}} \big(C_{_{[3]}}\big(\theta^{1},\theta^{2}, \dots, \theta^{N}\big)\big)}{C_{_{[3]}}\big(\theta^{1},\theta^{2}, \dots, \theta^{N}\big)}v
=\prod_{i=1}^{N}\lambda_i^2\frac{C_{_{[3]}}\big(\b\theta^{1},\b\theta^{2}, \dots, \b\theta^{N}\big)}{T_{_{n_3}} \big(C_{_{[3]}}\big(\b\theta^{1},\b\theta^{2}, \dots, \b\theta^{N}\big)\big)}v,
\end{gather}
\end{subequations}
leaves \eqref{matrix-LP} invariant. Otherwise, $\t{\b\phi}=\lambda^{-1}T_{_{n_3}} (\t\phi)$, $\t{\phi}=\lambda^{-1}T_{_{n_3}} \big(\t{\b\phi}\big)$.
\end{Proposition}

The d-p-mKdV equation \eqref{d-p-mKdV} is invariant with respect to the reversal of all lattice directions $n_i\rightarrow -n_i$, $i=1,2$. But its linear system~\eqref{matrix-LP} does not have such invariance and so the ref\/lections $n_i\rightarrow -n_i$, $i=1,2$, acting on~\eqref{matrix-LP} give a second linear system on the vector eigenfunction $\bm\Psi=(\psi,\b \psi)^T$, which also satisfy 2-periodic reduction condition
\begin{gather}\label{psi-condition}
\psi=\lambda^{-2}T_{_{n_3}}^{-2}(\psi), \qquad \b\psi=\lambda^{-1}T_{{n_3}}^{-1}(\psi),
\end{gather}
 as follows
\begin{subequations}\label{matrix-LP-2+}
\begin{gather}
\bm\Psi_{\b1}=\bm U\bm\Psi,\label{matrix-LP-2-1}\\
\bm\Psi_{\b2}=\bm V\bm\Psi, \label{matrix-LP-2-2}
\end{gather}
\end{subequations}
where
\begin{gather*}
\bm U=\left(
 \begin{matrix}
 a_1v_{\b1}v^{-1} & \lambda \\
 \lambda & a_1v_{\b 1}^{-1}v
 \end{matrix}
 \right)
,\qquad
\bm V=\left(
 \begin{matrix}
 a_2v_{\b2}v^{-1} & \lambda \\
 \lambda & a_2v_{\b2}^{-1}v
 \end{matrix}
 \right).
\end{gather*}
One then f\/inds that
\begin{gather*}
0=\bm\Psi_{\b1\b2}-\bm\Psi_{\b2\b1}=(\bm U_{\b2}\bm V-\bm V_{\b1}\bm U)\bm\Psi=\lambda Q\big(v,v_{\b1},v_{\b2},v_{\b{12}};a_1,a_2\big)
\left(
\begin{matrix}
 0 & -v_{\b1}^{-1}v_{\b2}^{-1} \\
v^{-1} v_{\b1\b2}^{-1} & 0
\end{matrix}
\right)\bm\Psi.
\end{gather*}
Now we apply the ref\/lections $n_i \rightarrow -n_i$, $i=1, 2$, in order to deduce Darboux transformation for the second linear system as below.
\begin{Proposition}\label{prop2-dpmKdV}
Suppose $(\rho,\b\rho)^T$, which holds the $2$-periodic property $\rho=\mu^{-2}T_{_{n_3}}^{-2}(\rho)$, $\b\rho=\mu^{-1}T_{_{n_3}}^{-1}(\rho)$, is a vector solution of the linear system~\eqref{matrix-LP-2+} by taking $\lambda=\mu$ for some $v$, then
\begin{gather*}
\mathrm {DT}^{\rho, \b\rho}\colon \
\psi\rightarrow\frac{C_{_{[\b 3]}}(\rho, \psi)} {\rho},\qquad \b\psi\rightarrow\frac{C_{_{[\b 3]}}(\b\rho, \b\psi)} {\b\rho} ,\qquad v\rightarrow \frac{T_{_{n_3}}^{-1}(\rho)}{\rho}v=\mu^2\frac{\b\rho}{T_{_{n_3}}^{-1}(\b\rho)}v
\end{gather*}
leaves \eqref{matrix-LP-2+} invariant. Otherwise,
\begin{gather*}
\frac{C_{_{[\b 3]}}(\b\rho, \b\psi)} {\b\rho}=\lambda^{-1}T_{_{n_3}}^{-1} \left(\frac{C_{_{[\b 3]}}(\rho, \psi)} {\rho}\right),
\qquad \frac{C_{_{[\b 3]}}(\rho, \psi)} {\rho}=\lambda^{-1}T_{_{n_3}}^{-1} \left(\frac{C_{_{[\b 3]}}(\b\rho, \b\psi)} {\b\rho}\right).
\end{gather*}
\end{Proposition}

Next we write down the closed form expression for the result of $N$ applications of the above Darboux transformation, which give solutions in Casoratian determinant form.

\begin{Proposition}\label{prop2N-dpmKdV}
Let $\big(\rho^{1}, \b\rho^{1}\big)^T, \big(\rho^{2},\b\rho^{2}\big)^T, \dots, \big(\rho^{N}, \b\rho^{N}\big)^T$, satisfying $\rho^{k} = \lambda_k^{-2}T_{_{n_3}}^{-2}(\rho^{k})$, $\b\rho^{k} = \lambda_k^{-1}T_{_{n_3}}^{-1} (\rho^{k})$, be $N$ non-zero independent vector solutions of the linear system~\eqref{matrix-LP-2+} by taking $\lambda=\lambda_k$, $k=1, 2, \dots, N$, for some $v$, such that $C_{_{[\b 3]}}(\rho^{1},\rho^{2}, \dots, \rho^{N})\neq0$. Then
\begin{gather*}
\psi\rightarrow\t\psi=\frac{C_{_{[\b 3]}}\big(\rho^{1},\rho^{2}, \dots, \rho^{N},\psi\big)}{C_{_{[\b 3]}}\big(\rho^{1},\rho^{2}, \dots, \rho^{N}\big)},
\qquad \b\psi\rightarrow\t{\b\psi}=\frac{C_{_{[\b 3]}}\big(\b\rho^{1},\b\rho^{2}, \dots, \b\rho^{N},\b\psi\big)}{C_{_{[\b 3]}}\big(\b\rho^{1},\b\rho^{2}, \dots, \b\rho^{N}\big)},\\
 v\rightarrow \t v=\frac{T_{_{n_3}}^{-1} \big(C_{_{[\b 3]}}\big(\rho^{1},\rho^{2}, \dots, \rho^{N}\big)\big)}{C_{_{[\b 3]}}\big(\rho^{1},\rho^{2}, \dots, \rho^{N}\big)}v =\prod_{i=1}^{N}\lambda_i^2\frac{C_{_{[\b 3]}}\big(\b\rho^{1},\b\rho^{2}, \dots, \b\rho^{N}\big)}{T_{_{n_3}}^{-1} \big(C_{_{[\b 3]}}\big(\b\rho^{1}, \b\rho^{2}, \dots, \b\rho^{N}\big)\big)}v,
\end{gather*}
leaves \eqref{matrix-LP-2+} invariant. Otherwise, $\t{\b\psi}=\lambda^{-1}T_{_{n_3}}^{-1} (\t\psi )$, $\t{\psi}=\lambda^{-1}T_{_{n_3}}^{-1} \big(\t{\b\psi}\big)$.
\end{Proposition}

To construct a binary Darboux transformation, we introduce the potentials $\omega=\omega(\phi, \psi)$ and $\b\omega=\omega(\b\phi, \b\psi)$, def\/ined by the relations
\begin{subequations}\label{dpmKdV-omega}
\begin{gather}
\Delta_3(\omega(\phi, \psi)) =\phi T_{_{n_3}} (\psi), \label{d-p-mKdV-omega-1}\\
\Delta_3(\omega(\b\phi, \b\psi)) =\b\phi T_{_{n_3}} (\b\psi). \label{d-p-mKdV-omega-2}
\end{gather}
\end{subequations}
If $(\phi, \b\phi)$ and $(\psi,\b\psi)$ satisfy the linear systems~\eqref{matrix-LP} and~\eqref{matrix-LP-2+} for some~$v$, respectively. Otherwise, together with the reductions~\eqref{phi-condition} and~\eqref{psi-condition}, we have the reduction condition for $(\omega,\b\omega)^T$ as follows
\begin{gather*}
T_{_{n_3}}(\omega(\phi, \psi))=\omega(\b\phi, \b\psi),\qquad T_{_{n_3}}^2(\omega(\phi, \psi))=\omega(\phi, \psi),\\
T_{_{n_3}}(\omega(\b\phi, \b\psi))=\omega(\phi, \psi),\qquad T_{_{n_3}}^2(\omega(\b\phi,\b\psi))=\omega(\b\phi, \b\psi).
\end{gather*}
Especially,
\begin{gather*}
T_{_{n_3}}(\omega(\phi, \rho))=\lambda\mu^{-1}\omega(\b\phi, \b\rho), \qquad T_{_{n_3}}^2(\omega(\phi, \rho))=\lambda^2\mu^{-2}\omega(\phi, \rho), \\
T_{_{n_3}}(\omega(\b\phi, \b\rho))=\lambda\mu^{-1}\omega(\phi, \rho), \qquad T_{_{n_3}}^2(\omega(\b\phi, \b\rho))=\lambda^2\mu^{-2}\omega(\b\phi, \b\rho);
\\
T_{_{n_3}}(\omega(\theta, \psi))=\lambda^{-1}\mu~\omega(\b\theta, \b\psi),\qquad T_{_{n_3}}^2(\omega(\theta, \psi))=\lambda^{-2}\mu^2\omega(\theta, \psi),\\
T_{_{n_3}}(\omega(\b\theta, \b\psi))=\lambda^{-1}\mu~\omega(\theta, \psi),\qquad T_{_{n_3}}^2(\omega(\b\theta, \b\psi))=\lambda^{-2}\mu^2\omega(\b\theta, \b\psi);
\\
T_{_{n_3}}(\omega(\theta, \rho))=\omega(\b\theta, \b\rho),\qquad T_{_{n_3}}^2(\omega(\theta, \rho))=\omega(\theta, \rho),\\
T_{_{n_3}}(\omega(\b\theta, \b\rho))=\omega(\theta, \rho),\qquad T_{_{n_3}}^2(\omega(\b\theta,\b\rho))=\omega(\b\theta, \b\rho);
\end{gather*}
and
\begin{gather*}
T_{_{n_3}}(\omega(\theta^k, \rho^l))=\left(\frac{\lambda_k}{\lambda_l}\right)\omega(\b\theta^k, \b\rho^l),\qquad T_{_{n_3}}^2(\omega(\theta^k, \rho^l))=\left(\frac{\lambda_k}{\lambda_l}\right)^2\omega(\theta^k, \rho^l),\\
T_{_{n_3}}(\omega(\b\theta^k, \b\rho^l))=\left(\frac{\lambda_k}{\lambda_l}\right)\omega(\theta^k, \rho^l),\qquad T_{_{n_3}}^2(\omega(\b\theta^k, \b\rho^l))=\left(\frac{\lambda_k}{\lambda_l}\right)^2\omega(\b\theta^k, \b\rho^l).
\end{gather*}
The following proposition gives the binary Darboux transformation of the d-p-mKdV equation.

\begin{Proposition}\label{binary-dpmKdV}
For some $v$, let $(\theta,\b\theta)^T$ and $(\phi,\b\phi)^T$ be two non-zero vector solutions of the linear system \eqref{matrix-LP}, respectively corresponding to spectrum parameters~$\mu$ and~$\lambda$; $(\rho,\b\rho)^T$ and $(\psi, \b\psi)^T$ be two non-zero vector solutions of the linear system~\eqref{matrix-LP-2+}, respectively corresponding to spectrum parameters~$\mu$ and~$\lambda$, then
\begin{alignat}{3}
& \mathrm{BDT}\colon \quad && \phi \rightarrow \phi-\theta \omega(\theta, \rho)^{-1}\omega(\phi,\rho),~\b\phi \rightarrow \b\phi-\b\theta \omega(\b\theta, \b\rho)^{-1}\omega(\b\phi,\b\rho),& \nonumber\\
&&& v \rightarrow \frac{T_{_{n_3}} (\omega(\theta, \rho))}{\omega(\theta, \rho)}v = \frac{\omega(\b\theta, \b\rho)}{T_{_{n_3}} (\omega(\b\theta, \b\rho))}v,& \label{dpmKdV-bDT-1}\\
& \mathrm{aBDT}\colon \quad && \psi \rightarrow \psi-\rho \omega(\theta, \rho)^{-1}\omega(\theta,\psi),~\b\psi \rightarrow \b\psi-\b\rho\omega(\b\theta, \b\rho)^{-1}\omega(\b\theta,\b\psi),&\nonumber\\
&&& v \rightarrow \frac{T_{_{n_3}} (\omega(\theta, \rho))}{\omega(\theta, \rho)}v = \frac{\omega(\b\theta, \b\rho)}{T_{_{n_3}} (\omega(\b\theta, \b\rho))}v, & \label{dpmKdV-bDT-2}
\end{alignat}
leave \eqref{matrix-LP} and \eqref{matrix-LP-2+} respectively invariant. Otherwise,
\begin{gather*}
\b\phi-\b\theta \omega(\b\theta, \b\rho)^{-1}\omega(\b\phi,\b\rho)=\lambda^{-1}T_{_{n_3}}\big(\phi- \theta \omega(\theta, \rho)^{-1}\omega(\phi,\rho)\big),\\
\phi-\theta \omega(\theta, \rho)^{-1} \omega(\phi, \rho) = \lambda^{-1}T_{_{n_3}} \big(\b\phi - \b\theta \omega(\b\theta, \b\rho)^{-1} \omega(\b\phi,\b\rho)\big),\\
 \b\psi - \b\rho \omega(\b\theta, \b\rho)^{-1}\omega(\b\theta,\b\psi) = \lambda^{-1}T_{_{n_3}}^{-1} \big(\psi - \rho \omega(\theta, \rho)^{-1}\omega(\theta,\psi)\big),\\
 \psi - \rho \omega(\theta, \rho)^{-1}\omega(\theta,\psi) = \lambda^{-1}T_{_{n_3}}^{-1} \big(\b\psi - \b\rho \omega(\b\theta, \b\rho)^{-1}\omega(\b\theta,\b\psi)\big).
\end{gather*}
\end{Proposition}

The $N$-fold iteration of these binary Darboux transformations are given below.

\begin{Proposition}\label{N-binary-dpmKdV}
Let $\big(\theta^{1} , \b\theta^{1}\big)^T , \big(\theta^{2} ,\b\theta^{2}\big)^T \!, \dots, \big(\theta^{N} , \b\theta^{N}\big)^T $ and $\big(\rho^{1} , \b\rho^{1}\big)^T , \big(\rho^{2} ,\b\rho^{2}\big)^T \!, \dots, \big(\rho^{N} , \b\rho^{N}\big)^T $ be $N$ independent vector solutions, holding $\theta^{k} = \lambda_k^{-2}T_{_{n_3}}^2(\theta^{k})$, $\b\theta^{k} = \lambda_k^{-1}T_{_{n_3}}(\theta^{k})$, and $\rho^{k} = \lambda_k^{-2}T_{_{n_3}}^{-2}(\rho^{k})$, $\b\rho^{k} = \lambda_k^{-1}T_{_{n_3}}^{-1}(\rho^{k})$, by taking $\lambda=\lambda_k$, $k=1, 2, \dots, N$, satisfy linear systems~\eqref{matrix-LP} and~\eqref{matrix-LP-2+} for some~$v$ respectively. Then
\begin{gather*}
\phi \rightarrow \h\phi = \frac{
\begin{vmatrix}
\omega\big(\bm\theta, \bm \rho^T \big) & \bm\theta\\
\omega\big(\phi, \bm \rho^T\big) & \phi
\end{vmatrix}}
{\big|\omega\big(\bm\theta, \bm \rho^T \big) \big|} ,
\qquad
\b\phi \rightarrow \h{\b\phi} = \frac{
\begin{vmatrix}
\omega\big(\bm{\b\theta}, \bm{\b \rho}^T \big) & \bm{\b\theta}\\
\omega\big(\b\phi, \bm {\b\rho}^T \big) & \b\phi
\end{vmatrix}}
{\big|\omega\big(\bm{\b\theta} , \bm {\b\rho^T}\big)\big |},
\\
v \rightarrow \h v = \frac{\big|T_{_{n_3}} \big(\omega(\bm\theta, \bm \rho^T )\big) \big|}{\big|\omega\big(\bm\theta, \bm \rho^T \big) \big|}v
 = \frac{\big|\omega\big(\bm{\b\theta} , \bm {\b\rho}^T \big) \big|}{\big|T_{_{n_3}} \big(\omega\big(\bm{\b\theta} , \bm {\b\rho}^T \big)\big) \big|}v,
\end{gather*}
 and
\begin{gather*}
\psi \rightarrow \h\psi =
\frac{
\begin{vmatrix}
\omega\big(\bm\theta^T, \bm \rho\big) & \bm\rho\\
\omega\big(\bm\theta^T, \psi\big) & \psi
\end{vmatrix}}
{\big|\omega\big(\bm\theta^T , \bm \rho\big) \big|},
\qquad
\b\psi \rightarrow \h{\b\psi} = \frac{
\begin{vmatrix}
\omega\big(\bm{\b\theta}^T, \bm{\b \rho}\big) & \bm{\b\rho}\\
\omega\big(\bm {\b\theta}^T,\b\psi\big) & \b\psi
\end{vmatrix}}
{\big|\omega\big(\bm{\b\theta}^T , \bm {\b\rho}\big) \big|},
\\
v \rightarrow \h v = \frac{\big|T_{_{n_3}} \big(\omega\big(\bm\theta^T , \bm \rho\big)\big) \big|}{\big|\omega\big(\bm\theta^T , \bm \rho\big)\big|}v
 = \frac{\big|\omega\big(\bm{\b\theta}^T , \bm {\b\rho} \big) \big|}{\big|T_{_{n_3}} \big(\omega\big(\bm{\b\theta}^T , \bm {\b\rho} \big)\big) \big|}v,
\end{gather*}
leave \eqref{matrix-LP} and \eqref{matrix-LP-2+} respectively invariant, where $\bm\theta=\big(\theta^1,\dots,\theta^N\big)^T$ and $\bm\rho=\big(\rho^1,\dots,\rho^N\big)^T$. Otherwise, $\h{\b\phi}=\lambda^{-1}T_{_{n_3}}(\h\phi)$, $\h{\phi}=\lambda^{-1}T_{_{n_3}}(\h{\b\phi})$, $\h{\b\psi}=\lambda^{-1}T_{_{n_3}}^{-1}(\h\psi)$, $\h{\psi}=\lambda^{-1}T_{_{n_3}}^{-1}(\h{\b\psi})$.
\end{Proposition}

\subsection{Explicit solutions obtained by Darboux transformations}
Here we present explicit examples of the classes of solutions that may be obtained by means of the Darboux transformations derived above.
We choose the seed solution of the d-p-mKdV equation~\eqref{d-p-mKdV} as $v=v_0=1$. With this choice, the f\/irst linear system \eqref{matrix-LP} reads
\begin{gather*}
\phi_1=a_1\phi+\lambda\b\phi,\qquad
\b\phi_1=a_1\b\phi+\lambda\phi,\\
\phi_2=a_2\phi+\lambda\b\phi,\qquad
\b\phi_2=a_2\b\phi+\lambda\phi,
\end{gather*}
and the eigenfunctions are found to be
\begin{subequations}\label{dpmKdV-seed-linear-solu-1}
\begin{gather}
\phi(n_1, n_2, n_3; \lambda)=\lambda^{n_3}\prod_{i=1}^{2}(a_i+\lambda)^{n_i}+(-\lambda)^{n_3}\prod_{i=1}^{2}(a_i-\lambda)^{n_i},\\
\b\phi(n_1, n_2, n_3; \lambda)=\lambda^{n_3}\prod_{i=1}^{2}(a_i+\lambda)^{n_i}-(-\lambda)^{n_3}\prod_{i=1}^{2}(a_i-\lambda)^{n_i},
\end{gather}
\end{subequations}
which hold $\phi=\lambda^{-2}T_{_{n_3}}(\phi)$, $\b\phi=\lambda^{-1}T_{_{n_3}}(\phi)$.

In a similar way the eigenfunctions of the second linear system \eqref{matrix-LP-2+} are
 \begin{subequations}\label{dpmKdV-seed-linear-solu-2}
\begin{gather}
\psi(n_1, n_2, n_3; \lambda)=\lambda^{-n_3}\prod_{i=1}^{2}(a_i+\lambda)^{-n_i}+(-\lambda)^{-n_3}\prod_{i=1}^{2}(a_i-\lambda)^{-n_i},\\
\b\psi(n_1, n_2, n_3; \lambda)=\lambda^{-n_3}\prod_{i=1}^{2}(a_i+\lambda)^{-n_i}-(-\lambda)^{-n_3}\prod_{i=1}^{2}(a_i-\lambda)^{-n_i},
\end{gather}
\end{subequations}
which hold $\psi=\lambda^{-2}T_{_{n_3}}^{-1}(\psi)$, $\b\psi=\lambda^{-1}T_{_{n_3}}^{-1}(\psi)$.

For these eigenfunctions \eqref{dpmKdV-seed-linear-solu-1} and \eqref{dpmKdV-seed-linear-solu-2} above we may integrate \eqref{dpmKdV-omega} and obtain the potentials
\begin{subequations}\label{dpmKdV-seed-linear-solu-omeg}
\begin{gather}
\omega(\phi, \psi)=\frac{1}{2\lambda}\left[(-1)^{n_3}\prod_{i=1}^{2}\left(\frac{a_i+\lambda}{a_i-\lambda}\right)^{n_i}
-(-1)^{-n_3}\prod_{i=1}^{2}\left(\frac{a_i+\lambda}{a_i-\lambda}\right)^{-n_i}\right],\\
\omega(\b\phi, \b\psi)=\frac{1}{2\lambda}\left[(-1)^{-n_3}\prod_{i=1}^{2}\left(\frac{a_i+\lambda}{a_i-\lambda}\right)^{-n_i}
-(-1)^{n_3}\prod_{i=1}^{2}\left(\frac{a_i+\lambda}{a_i-\lambda}\right)^{n_i}\right],
\end{gather}
\end{subequations}
which hold $\omega(\phi, \psi)=T_{_{n_3}}^2(\omega(\phi, \psi))$, $\omega(\b\phi, \b\psi)=T_{_{n_3}}(\omega(\phi, \psi))$.

Otherwise, for $\lambda=\lambda_k$, $v=v_0=1$, the f\/irst linear system \eqref{matrix-LP} has eigenfunctions
\begin{subequations}\label{dpmKdV-seed-linear-solu-theta}
\begin{gather}
\theta^k(n_1, n_2, n_3; \lambda_k)=\lambda_k^{n_3}\prod_{i=1}^{2}(a_i+\lambda_k)^{n_i}+(-\lambda_k)^{n_3}\prod_{i=1}^{2}(a_i-\lambda_k)^{n_i},\\
\b\theta^k(n_1, n_2, n_3; \lambda_k)=\lambda_k^{n_3}\prod_{i=1}^{2}(a_i+\lambda_k)^{n_i}-(-\lambda_k)^{n_3}\prod_{i=1}^{2}(a_i-\lambda_k)^{n_i},
\end{gather}
\end{subequations}
which hold $\theta^k=\lambda_k^{-2}T_{_{n_3}}^2(\theta^k)$, $\b\theta^k=\lambda_k^{-1}T_{_{n_3}}(\theta^k)$.

Similarly, for $\lambda=\lambda_l$, $v=v_0=1$, the second linear system~\eqref{matrix-LP-2+} has eigenfunctions
\begin{subequations}\label{dpmKdV-seed-linear-solu-rho}
\begin{gather}
\rho^l(n_1, n_2, n_3; \lambda_l)=\lambda_l^{-n_3}\prod_{i=1}^{2}(a_i+\lambda_l)^{-n_i}+(-\lambda_l)^{-n_3}\prod_{i=1}^{2}(a_i-\lambda_l)^{-n_i},\\
\b\rho^l(n_1, n_2, n_3; \lambda_l)=\lambda_l^{-n_3}\prod_{i=1}^{2}(a_i+\lambda_l)^{-n_i}-(-\lambda_l)^{-n_3}\prod_{i=1}^{2}(a_i-\lambda_l)^{-n_i},
\end{gather}
\end{subequations}
which hold $\rho^l=\lambda_l^{-2}T_{_{n_3}}^{-2}(\rho^l)$, $\b\rho^l=\lambda_l^{-1}T_{_{n_3}}^{-1}(\rho^l)$.

For these eigenfunctions \eqref{dpmKdV-seed-linear-solu-theta} and~\eqref{dpmKdV-seed-linear-solu-rho} above we may integrate~\eqref{dpmKdV-omega} and obtain the potential, for $\lambda_k\neq\lambda_l$,
\begin{subequations}\label{dpmKdV-seed-linear-solu-omeg-theta-rho}\allowdisplaybreaks
\begin{gather}
 \omega\big(\theta^k, \rho^l\big) = \lambda_l^{ -1} \left(\frac{\lambda_k}{\lambda_l}\right)^{n_3} \Bigg[
 \left(\frac{\lambda_k}{\lambda_l} - 1\right)^{ -1} \prod_{i=1}^{2} \left(\frac{a_i+\lambda_k}{a_i+\lambda_l}\right)^{n_i}\nonumber\\
\hphantom{\omega\big(\theta^k, \rho^l\big) =}{}
 - \left(\frac{\lambda_k}{\lambda_l} + 1\right)^{ -1} (-1)^{n_3} \prod_{i=1}^{2}\left(\frac{a_i-\lambda_k}{a_i+\lambda_l}\right)^{n_i}
+\left(\frac{\lambda_k}{\lambda_l} + 1\right)^{ -1} (-1)^{n_3} \prod_{i=1}^{2}\left(\frac{a_i+\lambda_k}{a_i-\lambda_l}\right)^{n_i}\nonumber\\
\hphantom{\omega\big(\theta^k, \rho^l\big) =}{}
- \left(\frac{\lambda_k}{\lambda_l} - 1\right)^{ -1} \prod_{i=1}^{2} \left(\frac{a_i-\lambda_k}{a_i-\lambda_l}\right)^{n_i}\Bigg],
\\
 \omega\big(\b\theta^k, \b\rho^l\big) = \lambda_l^{ -1} \left(\frac{\lambda_k}{\lambda_l}\right)^{n_3} \Bigg[
\left(\frac{\lambda_k}{\lambda_l} - 1\right)^{ -1} \prod_{i=1}^{2} \left(\frac{a_i+\lambda_k}{a_i+\lambda_l}\right)^{n_i}\nonumber\\
\hphantom{\omega\big(\b\theta^k, \b\rho^l\big) =}{}
 + \left(\frac{\lambda_k}{\lambda_l} + 1\right)^{ -1} (-1)^{n_3} \prod_{i=1}^{2}\left(\frac{a_i-\lambda_k}{a_i+\lambda_l}\right)^{n_i} -\left(\frac{\lambda_k}{\lambda_l} + 1\right)^{ -1} (-1)^{n_3} \prod_{i=1}^{2}\left(\frac{a_i+\lambda_k}{a_i-\lambda_l}\right)^{n_i} \nonumber\\
\hphantom{\omega\big(\b\theta^k, \b\rho^l\big) =}{}
- \left(\frac{\lambda_k}{\lambda_l} - 1\right)^{ -1} \prod_{i=1}^{2} \left(\frac{a_i-\lambda_k}{a_i-\lambda_l}\right)^{n_i}
\Bigg],
\end{gather}
\end{subequations}
which hold
\begin{gather*}
T_{_{n_3}}\big(\omega\big(\theta^k, \rho^l\big)\big)=\left(\frac{\lambda_k}{\lambda_l}\right)\omega\big(\b\theta^k, \b\rho^l\big), \qquad T_{_{n_3}}^2\big(\omega\big(\theta^k, \rho^l\big)\big)=\left(\frac{\lambda_k}{\lambda_l}\right)^2\omega\big(\theta^k, \rho^l\big).
\end{gather*}
For $\lambda_k=\lambda_l$, these eigenfunctions are~\eqref{dpmKdV-seed-linear-solu-omeg} taking $\lambda=\lambda_k=\lambda_l$.

Given the above expression it is straightforward to write down the following explicit solution for the d-p-mKdV equation \eqref{d-p-mKdV}
\begin{gather*}
v(n_1, n_2)=\frac{T_{_{n_3}} \big(C_{_{[3]}}\big(\theta^{1},\theta^{2}, \dots, \theta^{N}\big)\big)}{C_{_{[3]}}\big(\theta^{1},\theta^{2}, \dots, \theta^{N}\big)}v_0,
\end{gather*}
where $\theta^k=\theta^k(n_1, n_2, n_3;\lambda_k)$ is given by \eqref{dpmKdV-seed-linear-solu-theta} and $\lambda_k$ are arbitrary constants;
\begin{gather*}
v(n_1, n_2)=\frac{T_{_{n_3}}^{-1} \big(C_{_{[\b 3]}}\big(\rho^{1},\rho^{2}, \dots, \rho^{N}\big)\big)}{C_{_{[\b 3]}}\big(\rho^{1}, \rho^{2}, \dots, \rho^{N}\big)}v_0,
\end{gather*}
where $\rho^k=\rho^k(n_1, n_2, n_3;\lambda_k)$ is given by \eqref{dpmKdV-seed-linear-solu-rho} and $\lambda_k$ are arbitrary constants;
\begin{gather*}
v(n_1, n_2, n_3)=\frac{T_{_{n_3}} \left(\det (\omega_{k,l})\right)}{\det (\omega_{k,l})}v_0,\qquad k, l= 1, 2, \dots, N,
\end{gather*}
where $\omega_{k,l}$ is given by \eqref{dpmKdV-seed-linear-solu-omeg-theta-rho} with $\omega_{k,l}=\omega(\theta^k, \rho^l)$.

\section{Conclusions}
In this paper, we presents two main results. In the f\/irst we show how the d-p-mKdV equation and its Lax pairs in matrix form arise from the Hirota--Miwa equation by 2-periodic reduction. The second is that Darboux transformations and binary Darboux transformations are derived for the d-p-mKdV equation and we show how these may be used to construct exact solutions. In this paper, we have revisited the Darboux and binary transformations of the Hirota--Miwa equation but in a departure from the results in \cite{Nimmo-1997, Nimmo-Chaos-2000, Shi-2014}, by the gauge transformation \mbox{$\phi\rightarrow \prod\limits_{i=1}^{3}a_i^{-n_i}\phi$}, we write the linear system of Hirota--Miwa equation in a way which is suitable for obtaining the Lax pair of the d-pmKdV equation naturally by a 2-periodic reduction. Up to gauge transformations, these Lax pairs, which allow the application of the classical Darboux transformations, are coincident with the ones given by the multidimensional consistency property~\cite{BS-2008}. Hieta\-rin\-ta and Zhang~\cite{Hietarinta-2009} derived the $N$-soliton solutions to the d-p-mKdV equation using Hirota's direct method and the authors mention that the bilinear equations they get are similar to the Hirota--Miwa equation~\eqref{H-M-1}. The results in this paper, in which similar results are obtained by reduction of the Hirota--Miwa equation, give an explanation of the observations in~\cite{Hietarinta-2009}.

\appendix
\section{Some proofs}\allowdisplaybreaks
 This section contains proofs of some of the propositions in the main text. For the d-p-mKdV equation, one each of the $N$-fold basic Darboux transformations and binary Darboux transformations is proved. The omitted proofs are very similar.
\subsection{Proof of Proposition~\ref{prop1N-dpmKdV}}
Let
\begin{gather*}
F:= C_{_{[3]}}\big(\theta^1, \theta^2, \dots, \theta^N\big) = |\bm\theta(0),\bm\theta(1), \dots, \bm\theta(N-1)|, \\
G:= C_{_{[3]}}\big(\theta^1, \theta^2, \dots, \theta^N,\phi\big) = \big|\bm\theta^{^\dag} (0),\bm\theta^{^\dag} (1), \dots, \bm\theta^{^\dag} (N)\big|,\\
\b F:= C_{_{[3]}}\big(\b\theta^1, \b\theta^2, \dots, \b\theta^N\big) = \big|\b{\bm\theta}(0), \b{\bm\theta}(1), \dots, \b{\bm\theta}(N-1)\big|,\\
\b G:= C_{_{[3]}}\big(\b\theta^1,\b\theta^2, \dots, \b\theta^N,\phi\big) = \big|\b{\bm\theta}^{\dag} (0),\b{\bm\theta}^{\dag} (1), \dots,\b{\bm\theta}^{\dag} (N)\big|,
\end{gather*}
where $\bm\theta = \big(\theta^1, \theta^2, \dots, \theta^N\big)$, $\bm\theta^{^\dag}\! = \big(\theta^1, \theta^2, \dots, \theta^N \!,\phi\big)$, $\b{\bm\theta} = \big(\b\theta^1, \b\theta^2, \dots, \b\theta^N\big)$, $\b{\bm\theta}^{^\dag}\!=\big(\b\theta^1, \b\theta^2, \dots, \b\theta^N\! ,\b\phi\big)$. Moreover, by the reduction conditions $\b\phi=\lambda^{-1}T_{_{n_3}}(\phi)$, $\phi=\lambda^{-1}T_{_{n_3}}(\b\phi)$, $\b\theta^i=\lambda_i^{-1}T_{_{n_3}} (\theta^i)$, and $\theta^i=\lambda_i^{-1}T_{_{n_3}} (\b\theta^i)$, we easily get the relation $\frac{\b G}{\b F}=\lambda^{-1}T_{_{n_3}} \big(\frac{G}{F}\big)$, $\frac{ G}{F}=\lambda^{-1}T_{_{n_3}} \big(\frac{\b G}{\b F}\big)$. To verify that~\eqref{matrix-LP} is invariant under~\eqref{N-DT1} we must show that, for $\kappa=1, 2$,
\begin{gather*}
G_\kappa T_{_{n_3}} (F)= a_\kappa v_{\kappa}v^{-1}T_{_{n_3}} (F_\kappa)G+T_{_{n_3}} (G)F_\kappa,\\ 
\b G_\kappa T_{_{n_3}} (\b F)= a_\kappa v_{\kappa}^{-1} vT_{_{n_3}} (\b F_\kappa)\b G+T_{_{n_3}} (\b G)\b F_\kappa, 
\end{gather*}
which is equivalent to, under the backward shift operator $T_{_{n_3}}^{-1}T_{_{n_{_\kappa}}}^{-1}$,
\begin{subequations}
\begin{gather}
T_{_{n_3}}^{-1} (G) F_{_{\b\kappa}}= a_\kappa v v_{\b \kappa}^{-1}FT_{_{n_3}}^{-1} (G_{_{\b\kappa}})+G_{_{\b \kappa}}T_{_{n_3}}^{-1} (F), \label{bilinear-form-1-a}\\
 T_{_{n_3}}^{-1} (\b G)\b F_{_{\b\kappa}}= a_\kappa v^{-1} v_{\b \kappa}\b FT_{_{n_3}}^{-1} (\b G_{_{\b\kappa}})+\b G_{_{\b \kappa}}T_{_{n_3}}^{-1} (\b F).\label{bilinear-form-1-b}
\end{gather}
\end{subequations}
From \eqref{matrix-LP}, we can deduce the following formulae which are the basic properties we use in proving
\begin{gather*}
\bm\theta_{\b\kappa}(l) = \bm\theta(l-1)+\sum_{i=0}^{l-2}(-\alpha_{\b\kappa})^{i+1}\bm\theta(l-2-i)+(-\alpha_{\b\kappa})^l \bm\theta_{\b\kappa}(0),\\ 
{\bm\theta^\dag}_{ \b\kappa}(l) = \bm\theta^\dag(l-1)+\sum_{i=0}^{l-2}(-\alpha_{\b\kappa})^{i+1}{\bm\theta^\dag}(l-2-i)+(-\alpha_{\b\kappa})^l {\bm\theta^\dag}_{ \b\kappa}(0), \\ 
\bm{\b\theta}_{\b\kappa}(l) = \bm{\b\theta}(l-1)+\sum_{i=0}^{l-2}(-\beta_{\b\kappa})^{i+1}\bm{\b\theta}(l-2-i)+(-\beta_{\b\kappa})^l \bm{\b\theta}{_{_{\b\kappa}}}(0),\\ 
{\bm{\b\theta}^\dag} _{ \b\kappa}(l) = \bm{\b\theta}^\dag(l-1)+\sum_{i=0}^{l-2}(-\beta_{\kappa})^{i+1}{\bm{\b\theta}^\dag}(l-2-i)+(-\beta_{\b\kappa})^l {\bm{\b\theta}^\dag}_{ \b\kappa}(0), 
\end{gather*}
where $\alpha_{\b\kappa}=a_\kappa v v_{\b\kappa}^{-1}$ and $\beta_{\b\kappa}=a_\kappa v^{-1} v_{\b\kappa}$ are scalars, $\kappa=1, 2$. Then, for \eqref{bilinear-form-1-a}, it follows that
\begin{gather*}
T_{_{n_3}}^{-1}(G)= \big|\bm {\theta^\dag} (-1), \bm{\theta^\dag} (0), \bm{\theta^\dag} (1), \dots, \bm{\theta^\dag} (N-1)\big|, \\ 
F_{_{\b\kappa}}= \big|\bm {\theta_{ \b\kappa}^\dag}(0), \bm{\theta}(0), \bm{\theta}(1), \dots, \bm{\theta}(N-2)\big|, \\ 
\alpha_{\kappa}T_{_{n_3}}^{-1} (G_{_{\b\kappa}})= -\big|\bm {\theta^\dag}_{\b\kappa} (0), \bm{\theta^\dag} (-1), \bm{\theta^\dag} (0), \dots, \bm{\theta^\dag} (N-2)\big|,\\ 
F= \big|\bm{\theta}(0), \bm{\theta}(1), \bm{\theta}(2), \dots, \bm{\theta}(N-1)\big|, \\ 
G_{_{\b \kappa}}= \big|\bm {\theta_{\b\kappa}^\dag} (0), \bm{\theta^\dag} (0), \bm{\theta^\dag} (1), \dots, \bm{\theta^\dag} (N-1)\big|,\\ 
T_{_{n_3}}^{-1} (F)= \big|\bm{\theta}(-1), \bm{\theta}(0), \bm{\theta}(1), \dots, \bm{\theta}(N-2)\big|. 
\end{gather*}
Substituting into the left-hand side of \eqref{bilinear-form-1-a}, and using the Laplace theorem, we get
\begin{gather*}
 \text{LHS} = \left |
\begin{matrix}
\bm {\theta^\dag}_{\b\kappa}(0) & \bm{\theta^\dag}(-1) & \bm{\theta^\dag}(0) & \cdots & \bm{\theta^\dag}(N-2) & \bm0 & \cdots & \bm0 & \bm{\theta^\dag}(N-1)
\\
\bm {\theta}_{\b\kappa} (0) & \bm{\theta}(-1) & \bm0 & \cdots & \bm0 & \bm{\theta}(0) & \cdots & \bm{\theta}(N-2) & \bm{\theta}(N-1)
\end{matrix}
 \right | = 0. 
\end{gather*}
In a similar way, we can prove that \eqref{bilinear-form-1-b} is also satisf\/ied.

\subsection{Proof of Proposition~\ref{N-binary-dpmKdV}}
The proof is by induction. Let $\big(\theta^{1}, \b\theta^{1}\big)^T= (\theta[0], \b\theta[0] )^T, \big(\theta^{2} ,\b\theta^{2}\big)^T, \dots, \big(\theta^{N}, \b\theta^{N}\big)^T$ and $\big(\rho^{1}, \b\rho^{1}\big)^T=(\rho[0], \b\rho[0])^T, \big(\rho^{2},\b\rho^{2}\big)^T, \dots, \big(\rho^{N} , \b\rho^{N}\big)^T $ be vector eigenfunctions of `seed' linear system \eqref{matrix-LP} and \eqref{matrix-LP-2+} respectively for the `seed' potential $v=v[0]$ and let $(\phi, \b\phi)=(\phi[0], \b\phi[0])$, $(\psi, \b\psi)=(\psi[0], \b\psi[0])$ denote arbitrary eigenfunctions.

The $N$th iteration of binary Darboux transformations is via the formulae, $N=1, 2, \dots$,
\begin{subequations}\label{binary-N-DT}
\begin{gather}
 \phi[N]=\phi[N-1]-\theta[N-1]\omega(\theta[N-1], \rho[N-1])^{-1}\omega(\phi[N-1], \rho[N-1]), \label{A.2.1a}\\
 \b\phi[N]=\b\phi[N-1]-\b\theta[N-1]\omega(\b\theta[N-1], \b\rho[N-1])^{-1}\omega(\b\phi[N-1], \b\rho[N-1]), \label{A.2.1b}\\
 \psi[N]=\psi[N-1]-\rho[N-1]\omega(\theta[N-1], \rho[N-1])^{-1}\omega(\theta[N-1], \psi[N-1]), \label{A.2.1c}\\
 \b\psi[N]=\b\psi[N-1]-\b\rho[N-1]\omega(\b\theta[N-1], \b\rho[N-1])^{-1}\omega(\b\theta[N-1], \b\psi[N-1]), \label{A.2.1d}\\
 v[N]=\frac{T_{n_3}(\omega(\theta[N - 1], \rho[N - 1]))}{\omega(\theta[N - 1], \rho[N - 1])}v[N - 1]\nonumber\\
 \hphantom{v[N]}{}=\frac{\omega(\b\theta[N - 1], \b\rho[N - 1])}{T_{n_3}(\omega(\b\theta[N - 1], \b\rho[N - 1]))}v[N - 1],\label{A.2.1e}
\end{gather}
\end{subequations}
and
\begin{subequations}
\begin{gather}
\theta[N]=\phi[N]|_{_{\phi\rightarrow \theta^{N+1}}}, \qquad \b\theta[N]=\b\phi[N]|_{_{\b\phi \rightarrow \b\theta^{N+1}}}, \label{A.2.2a}\\
\rho[N]=\psi[N]|_{_{\psi\rightarrow \rho^{N+1}}}, \qquad \b\rho[N]=\b\psi[N]|_{_{\b\psi \rightarrow \b\rho^{N+1}}}. \label{A.2.2b}
\end{gather}
\end{subequations}

For $N=1$, the iterated binary Darboux transformation is the basic form given by~\eqref{dpmKdV-bDT-1} and~\eqref{dpmKdV-bDT-2}.

Suppose for $N=k$, the Proposition~\ref{N-binary-dpmKdV} is right, i.e., we have
\begin{subequations}\label{k-binary-dpmKdV-1}
\begin{gather}
\phi[k] = \begin{vmatrix}
\omega\big(\bm\theta, \bm \rho^T \big) & \bm\theta\\
\omega\big(\phi, \bm \rho^T\big) & \phi
\end{vmatrix}
 \big|\omega\big(\bm\theta, \bm \rho^T \big) \big|^{-1} ,\qquad
{\b\phi[k]} = \begin{vmatrix}
\omega\big(\bm{\b\theta}, \bm{\b \rho}^T \big) & \bm{\b\theta}\\
\omega\big(\b\phi, \bm {\b\rho}^T \big) & \b\phi
\end{vmatrix}
 \big|\omega\big(\bm{\b\theta} , \bm {\b\rho^T}\big) \big|^{-1} , \label{A.2.3a}\\
\psi[k] = \begin{vmatrix}
\omega\big(\bm\theta^T, \bm \rho\big) & \bm\rho\\
\omega\big(\bm\theta^T, \psi) & \psi
\end{vmatrix}
 \big|\omega\big(\bm\theta^T , \bm \rho\big) \big|^{-1},\qquad
{\b\psi[k]} = \begin{vmatrix}
\omega\big(\bm{\b\theta}^T, \bm{\b \rho}\big) & \bm{\b\rho}\\
\omega\big(\bm {\b\theta}^T,\b\psi\big) & \b\psi \\
\end{vmatrix}
 \big|\omega\big(\bm{\b\theta}^T , \bm {\b\rho}\big) \big|^{-1},\label{A.2.3b}\\
v[k] = \frac{\big|T_{_{n_3}} \big(\omega\big(\bm\theta^T , \bm \rho\big)\big) \big|}{\big|\omega\big(\bm\theta^T , \bm \rho\big)\big|}v
 = \frac{\big|\omega\big(\bm{\b\theta}^T , \bm {\b\rho} \big) \big|}{\big|T_{_{n_3}} \big(\omega\big(\bm{\b\theta}^T , \bm {\b\rho} \big)\big) \big|}v\nonumber\\
\hphantom{v[k]}{}
= \frac{\big|T_{_{n_3}} \big(\omega\big(\bm\theta , \bm \rho^T\big)\big) \big|}{\big|\omega\big(\bm\theta , \bm \rho^T\big)\big|}v = \frac{\big|\omega\big(\bm{\b\theta} , \bm {\b\rho^T} \big) \big|}{\big|T_{_{n_3}} \big(\omega\big(\bm{\b\theta} , \bm {\b\rho^T} \big)\big) \big|}v, \label{A.2.3c}
\end{gather}
\end{subequations}
where the column vectors are as below
\begin{gather*}
\bm\theta=\big(\theta^1, \theta^2, \dots, \theta^k\big)^T, \qquad \bm{\b\theta}=\big(\b\theta^1, \b\theta^2, \dots, \b\theta^k\big)^T, \\
\bm\rho=\big(\rho^1, \rho^2, \dots, \rho^k\big)^T, \qquad \bm{\b\rho}=\big(\b\rho^1, \b\rho^2, \dots, \b\rho^k\big)^T,
\end{gather*}
and the $k \times k$ matrices and $1 \times k$ row vectors are as follows
\begin{gather*}
\omega\big(\bm\theta, \bm\rho^T\big)
 = \left(\begin{matrix}
\omega\big(\theta^1, \rho^1\big) &\omega\big(\theta^1, \rho^2\big) & \cdots & \omega\big(\theta^1, \rho^k\big)\\
\omega\big(\theta^2, \rho^1\big) &\omega\big(\theta^2, \rho^2\big) & \cdots & \omega\big(\theta^2, \rho^k\big)\\
\vdots&\vdots & \cdots &\vdots\\
\omega\big(\theta^k, \rho^1\big) &\omega\big(\theta^k, \rho^2\big) & \cdots & \omega\big(\theta^k, \rho^k\big)
\end{matrix}
\right)
,\qquad \omega\big(\bm\theta^T, \bm\rho\big)=\omega\big(\bm\theta, \bm\rho^T\big)^T,
\\
\omega\big(\bm{\b\theta}, \bm{\b\rho}^T\big)
 = \left(\begin{matrix}
\omega\big(\b\theta^1, \b\rho^1\big) &\omega\big(\b\theta^1, \b\rho^2\big) & \cdots & \omega\big(\b\theta^1,\b \rho^k\big)\\
\omega\big(\b\theta^2,\b \rho^1\big) &\omega\big(\b\theta^2, \b\rho^2\big) & \cdots & \omega\big(\b\theta^2, \b\rho^k\big)\\
\vdots&\vdots & \cdots &\vdots\\
\omega\big(\b\theta^k, \b\rho^1\big) &\omega\big(\b\theta^k, \b\rho^2\big) & \cdots & \omega\big(\b\theta^k, \b\rho^k\big)
\end{matrix}
\right),\qquad
\omega\big(\bm{\b\theta}^T, \bm{\b\rho}\big)=\omega\big(\bm{\b\theta}, \bm{\b\rho}^T\big)^T,
\\
 \omega\big(\phi, \bm\rho^T\big)
=\left(\begin{matrix}
\omega\big(\phi, \rho^1\big) &\omega\big(\phi, \rho^2\big) & \cdots & \omega\big(\phi, \rho^k\big)
\end{matrix}
\right)
,\\
\omega\big(\b\phi, \bm{\b\rho}^T\big)
=\left(\begin{matrix}
\omega\big(\b\phi, \b\rho^1\big) &\omega\big(\b\phi, \b\rho^2\big) & \cdots & \omega\big(\b\phi, \b\rho^k\big)
\end{matrix}
\right),\\
 \omega\big(\bm\theta^T, \psi \big)
=\left(\begin{matrix}
\omega\big(\theta^1, \psi\big) &\omega\big(\theta^2, \psi\big) & \cdots & \omega\big(\theta^k, \psi\big)
\end{matrix}
\right),\\
\omega\big(\bm{\b\theta}^T, \b\psi \big)
=\left(\begin{matrix}
\omega\big(\b\theta^1, \b\psi\big) &\omega\big(\b\theta^2, \b\psi\big) & \cdots & \omega\big(\b\theta^k, \b\psi\big)
\end{matrix}
\right).
\end{gather*}
Moerover, we have
\begin{subequations}\label{omega-formula}
\begin{gather}
\omega(\phi[k], \psi[k]) =
\begin{vmatrix}
\omega\big(\bm\theta, \bm\rho^T\big) & \omega(\bm\theta, \psi)\\
\omega\big(\phi, \bm\rho^T\big) & \omega(\phi, \psi)
\end{vmatrix}
\big|\omega\big(\bm\theta, \bm\rho^T\big)\big|^{-1}, \label{A.2.4a}\\
\omega(\b\phi[k], \b\psi[k])=
\begin{vmatrix}
\omega\big(\bm{\b\theta}, \bm\rho^T\big) & \omega(\bm{\b\theta}, \b\psi)\\
\omega\big(\b\phi, \bm{\b\rho}^T\big) & \omega(\b\phi, \b\psi)
\end{vmatrix}
\big|\omega\big(\bm{\b\theta}, \bm{\b\rho}^T\big)\big|^{-1}. \label{A.2.4b}
\end{gather}
\end{subequations}
The proof of \eqref{omega-formula} is as follows. By the def\/inition of~$\omega$ and $\b\omega$ in~\eqref{dpmKdV-omega}, and together with~\eqref{A.2.3a} and~\eqref{A.2.3b}, we have
\begin{gather*}
\Delta_3 (\omega(\phi[k], \psi[k]) )= \phi[k]T_{_{n_3}}(\psi[k])
= \Delta_3 (\omega(\phi, \psi) )-T_{_{n_3}}\big(\omega\big(\bm\theta^T, \psi\big)\omega\big(\bm\theta^T, \bm\rho\big)^{-1}\big)\Delta_3(\omega(\phi, \bm\rho))\\
\hphantom{\Delta_3 (\omega(\phi[k], \psi[k]) )=}{}
-\Delta_3\big(\omega\big(\bm\theta^T, \psi\big)\big)\omega\big(\bm\theta^T, \bm\rho\big)^{-1}\omega(\phi, \bm\rho)\\
\hphantom{\Delta_3 (\omega(\phi[k], \psi[k]) )=}{}
+T_{_{n_3}}\big(\omega\big(\bm\theta^T, \psi\big)\omega\big(\bm\theta^T, \bm\rho\big)^{-1}\big)\Delta_3\big(\omega\big(\bm\theta^T, \bm\rho\big)\big)\omega\big(\bm\theta^T, \bm\rho\big)^{-1}\omega(\phi, \bm\rho) \\
\hphantom{\Delta_3 (\omega(\phi[k], \psi[k]) )}{}
=\Delta_3\big(\omega(\phi, \psi)-\omega\big(\bm\theta^T, \psi\big)\omega\big(\bm\theta^T, \bm\rho\big)^{-1}\omega(\phi, \bm\rho)\big)\\
\hphantom{\Delta_3 (\omega(\phi[k], \psi[k]) )}{}
= \Delta_3\left(\begin{vmatrix}
\omega\big(\bm\theta, \bm\rho^T\big) & \omega(\bm\theta, \psi)\\
\omega\big(\phi, \bm\rho^T\big) & \omega(\phi, \psi)
\end{vmatrix}
\big|\omega\big(\bm\theta, \bm\rho^T\big)\big|^{-1}\right),
\end{gather*}
which means \eqref{A.2.4a} is right. Similarly, we can get \eqref{A.2.4b} is right as well.

Note here that we use the dif\/ference operator property for matrices as follows
\begin{gather*}
\Delta_n\big(CA^{-1}B\big)=T_n(CA^{-1})\Delta_n(B)-\Delta_n(C)A^{-1}B-C_nA_n^{-1}\Delta_n(A)A^{-1}B,
\end{gather*}
where $\Delta_n=T_n-1$ is the dif\/ference operator, $A=A_{N\times N}$, $B=B_{N\times 1}$ and $C=C_{1\times N}$ are arbitrary function matrix, column vector and row vector of independent discrete variable~$n$ respectively.

Next, let us prove the $(k+1)$-th step is also right. By the iterated formulae~\eqref{A.2.1a}, we have
\begin{gather}
\phi[k+1]=\phi[k]-\theta[k]\omega(\theta[k], \rho[k])^{-1}\omega(\phi[k], \rho[k]). \label{A.2.5}
\end{gather}
Substitute \eqref{A.2.1a}, \eqref{A.2.2a}, \eqref{A.2.2b} and \eqref{A.2.4a} into \eqref{A.2.5}, we have
\begin{gather*}
 \phi[k + 1] =
\begin{vmatrix}
\omega\big(\bm\theta, \bm\rho^T\big) & \bm\theta\\
\omega\big(\phi, \bm\rho^T\big) & \phi\\
\end{vmatrix}
\big|\omega\big(\bm\theta, \bm\rho^T\big)\big|^{-1}-
\begin{vmatrix}
\omega\big(\bm\theta, \bm\rho^T\big) & \bm\theta\\
\omega\big(\theta^{k+1}, \bm\rho^T\big) & \theta^{k+1}
\end{vmatrix}
 \big|\omega\big(\bm\theta, \bm\rho^T\big)\big|^{-1} \\
\hphantom{\phi[k + 1] =}{} \times \begin{vmatrix}
\omega\big(\bm\theta, \bm\rho^T\big) & \omega\big(\bm\theta, \rho^{k+1}\big) \\
\omega\big(\theta^{k+1}, \bm\rho^T\big) & \omega\big(\theta^{k+1}, \rho^{k+1}\big)
\end{vmatrix}^{-1}
 \big|\omega\big(\bm\theta, \bm\rho^T\big)\big|\\
\hphantom{\phi[k + 1] =}{} \times
\begin{vmatrix}
\omega\big(\bm\theta, \bm\rho^T\big) & \omega\big(\bm\theta, \rho^{k+1}\big) \\
\omega\big(\phi, \bm\rho^T\big) & \omega\big(\phi, \rho^{k+1}\big)
\end{vmatrix}
\big|\omega\big(\bm\theta, \bm\rho^T\big)\big|^{-1}\\
\hphantom{\phi[k + 1] }{}
 = \left(\begin{vmatrix}
\omega\big(\bm\theta, \bm\rho^T\big) & \bm\theta\\
\omega\big(\phi, \bm\rho^T\big) & \phi
\end{vmatrix}
\begin{vmatrix}
\omega\big(\bm\theta, \bm\rho^T\big) & \omega\big(\bm\theta, \rho^{k+1}\big) \\
\omega\big(\theta^{k+1}, \bm\rho^T\big) & \omega\big(\theta^{k+1}, \rho^{k+1}\big)
\end{vmatrix}
 -
\begin{vmatrix}
\omega\big(\bm\theta, \bm\rho^T\big) & \bm\theta\\
\omega\big(\theta^{k+1}, \bm\rho^T\big) & \theta^{k+1}
\end{vmatrix}\right.\\
\left. \hphantom{\phi[k + 1] =}{}\times
\begin{vmatrix}
\omega\big(\bm\theta, \bm\rho^T\big) & \omega\big(\bm\theta, \rho^{k+1}\big) \\
\omega\big(\phi, \bm\rho^T\big) & \omega\big(\phi, \rho^{k+1}\big)
\end{vmatrix}
\right)\left(\begin{vmatrix}
\omega\big(\bm\theta, \bm\rho^T\big) & \omega\big(\bm\theta, \rho^{k+1}\big) \\
\omega\big(\theta^{k+1}, \bm\rho^T\big) & \omega\big(\theta^{k+1}, \rho^{k+1}\big) \\
\end{vmatrix} \big|\omega\big(\bm\theta, \bm\rho^T\big)\big|\right)^{-1}\\
\hphantom{\phi[k + 1] }{} =
\begin{vmatrix}
\omega\big(\bm\theta, \bm\rho^T\big) & \omega\big(\bm\theta, \rho^{k+1}\big) & \bm\theta\\
\omega\big(\theta^{k+1}, \bm\rho^T\big) & \omega\big(\theta^{k+1}, \rho^{k+1}\big) & \theta^{k+1}\\
\omega\big(\phi, \bm\rho^T\big) & \omega(\phi, \rho^{k+1}) & \phi
\end{vmatrix}
\begin{vmatrix}
\omega\big(\bm\theta, \bm\rho^T\big) & \omega\big(\bm\theta, \rho^{k+1}\big) \\
\omega\big(\theta^{k+1}, \bm\rho^T\big) & \omega\big(\theta^{k+1}, \rho^{k+1}\big)
\end{vmatrix}^{-1}.
\end{gather*}
Note here that we use the Jacobi identity.

Similarly, we can have
\begin{gather*}
\psi[k+1] =
\begin{vmatrix}
\omega\big(\bm\theta^T, \bm\rho\big) & \omega\big(\theta^{k+1}, \bm\rho\big) & \bm\rho\\
\omega\big(\bm\theta^T, \rho^{k+1}\big) & \omega\big(\theta^{k+1}, \rho^{k+1}\big) & \rho^{k+1}\\
\omega\big(\bm\theta^T, \psi\big) & \omega\big(\theta^{k+1}, \psi\big) & \psi
\end{vmatrix}
\begin{vmatrix}
\omega\big(\bm\theta^T, \bm\rho\big) & \omega\big(\theta^{k+1}, \bm\rho\big) \\
\omega\big(\bm\theta^T, \rho^{k+1}\big) & \omega\big(\theta^{k+1}, \rho^{k+1}\big)
\end{vmatrix}^{-1}.
\end{gather*}

Then from \eqref{A.2.1e}, for the potential $\h v$, we get
\begin{gather*}
v[k+1]= \frac{T_{n_3} (\omega(\theta[k], \rho[k]) )}{\omega(\theta[k], \rho[k])}v[k]
= T_{n_3} \left(
\begin{vmatrix}
\omega\big(\bm\theta, \bm\rho^T\big) & \omega(\bm\theta, \rho^{k+1})\\
\omega)\big(\theta^{k+1}, \bm\rho^T\big) & \omega\big(\theta^{k+1}, \rho^{k+1}\big)
\end{vmatrix}
\big|\omega\big(\bm\theta, \bm\rho^T\big)\big|^{-1}
\right) \\
 \hphantom{v[k+1]=}{} \times
\begin{vmatrix}
\omega\big(\bm\theta, \bm\rho^T\big) & \omega\big(\bm\theta, \rho^{k+1}\big)\\
\omega\big(\theta^{k+1}, \bm\rho^T\big) & \omega\big(\theta^{k+1}, \rho^{k+1}\big)
\end{vmatrix}^{-1}
\big|\omega\big(\bm\theta, \bm\rho^T\big)\big|
T_{n_3}\big(\big|
\omega\big(\bm\theta, \bm\rho^T\big)\big|\big) \big|\omega\big(\bm\theta, \bm\rho^T\big)\big|^{-1} v \\
\hphantom{v[k+1]}{} =
T_{n_3} \left(
\begin{vmatrix}
\omega\big(\bm\theta, \bm\rho^T\big) & \omega\big(\bm\theta, \rho^{k+1}\big)\\
\omega\big(\theta^{k+1}, \bm\rho^T\big) & \omega\big(\theta^{k+1}, \rho^{k+1}\big)
\end{vmatrix}
\right)
\begin{vmatrix}
\omega\big(\bm\theta, \bm\rho^T\big) & \omega\big(\bm\theta, \rho^{k+1}\big)\\
\omega \big(\theta^{k+1}, \bm\rho^T\big) & \omega\big(\theta^{k+1}, \rho^{k+1}\big)
\end{vmatrix}^{-1} v .
\end{gather*}

The proofs of the remaining parts are very similar, we omit them here.

\subsection*{Acknowledgements}
One of the author (YS) would like to acknowledge Professor Jarmo Hietarinta for his useful suggestions and hospitality when the author visited Turku University. The authors (YS and JXZ) also thank for the f\/inancial support from NSFC (Grant Numbers 11501510, 11271362, 11271266).

\pdfbookmark[1]{References}{ref}
\LastPageEnding

\end{document}